\documentclass[conference]{IEEEtran}
\IEEEoverridecommandlockouts
\usepackage{cite}
\usepackage{amsmath,amssymb,amsfonts}
\usepackage{algorithm}
\usepackage{algpseudocode}
\usepackage{graphicx}
\usepackage{textcomp}
\usepackage{xcolor}
\usepackage{url}
\usepackage{listings}
\usepackage{inconsolata} 
\usepackage{balance}

\def\BibTeX{{\rm B\kern-.05em{\sc i\kern-.025em b}\kern-.08em
    T\kern-.1667em\lower.7ex\hbox{E}\kern-.125emX}}

\newcommand{\lt}{\ensuremath <}
\newcommand{\gt}{\ensuremath >}
\begin{document}

\title{QoS-Aware Proportional Fairness Scheduling for Multi-Flow 5G UEs: A Smart Factory Perspective
\thanks{This publication has emanated from research conducted with the financial
support of Taighde Éireann – Research Ireland under Grant number 13/RC/2077 P2.
For the purpose of Open Access, the author has applied a CC-BY public
copyright license to any Author Accepted Manuscript version arising from
this submission}
}

\author{\IEEEauthorblockN{Mohamed Seliem, Utz Roedig, Cormac Sreenan, and Dirk Pesch}
\IEEEauthorblockA{\textit{School of Computer Science and Information Technology} \\
\textit{National Universities of Ireland, University College Cork}, Cork, Ireland \\
MSeliem@ucc.ie, u.roedig@ucc.ie, Cormac.Sreenan@ucc.ie, Dirk.pesch@ucc.ie}
}

\maketitle

\begin{abstract}
Private 5G networks are emerging as key enablers for smart factories, where a single device often handles multiple concurrent traffic flows with distinct Quality of Service (QoS) requirements. Existing simulation frameworks, however, lack the fidelity to model such multi-flow behavior at the QoS Flow Identifier (QFI) level. This paper addresses this gap by extending Simu5G to support per-QFI modeling and by introducing a novel QoS-aware Proportional Fairness (QoS-PF) scheduler. The scheduler dynamically balances delay, Guaranteed Bit Rate (GBR), and priority metrics to optimize resource allocation across heterogeneous flows. We evaluate the proposed approach in a realistic smart factory scenario featuring edge-hosted machine vision, real-time control loops, and bulk data transfer. Results show that QoS-PF improves deadline adherence and fairness without compromising throughput. All extensions are implemented in a modular and open-source manner to support future research. Our work provides both a methodological and architectural foundation for simulating and analyzing advanced QoS policies in industrial 5G deployments.
\end{abstract}

\begin{IEEEkeywords}
5G, QoS, scheduling, proportional fairness, QFI, industrial networks, Simu5G
\end{IEEEkeywords}

\section{Introduction: Can Private 5G Support the Smart Factory Dream?}
The promise of smart manufacturing hinges on the ability to coordinate diverse industrial applications—robotic control, visual inspection, telemetry, and autonomous logistics—within a unified, flexible, and low-latency communication fabric. As factories evolve toward Industry 5.0, wired solutions, though deterministic, are increasingly seen as rigid and costly to scale or reconfigure \cite{b1}. In response, private 5G networks have emerged as a transformative enabler, offering ultra-reliable wireless connectivity, time-sensitive communication, and flexible resource slicing \cite{b2, b3}.

Yet, realizing this vision in practice presents a formidable challenge: industrial traffic is inherently heterogeneous. A single UE (e.g., a robot controller or edge device) may carry simultaneous flows with sharply contrasting QoS demands—ranging from URLLC-grade control loops to high-throughput video streams and best-effort telemetry \cite{b4}. Each flow, identified via a QoS Flow Identifier (QFI), must receive service aligned with its 5QI-class properties, including guaranteed bit rate (GBR), delay bounds, and packet error rates \cite{b5}.

Current 5G simulators, however, are ill-equipped to explore this complexity. Most do not support multiple QFIs per UE, assume static or unrealistic traffic patterns, or rely on simplistic scheduling algorithms that overlook dynamic fairness and prioritization~\cite{b6, b7}. Consequently, researchers and practitioners lack the tools to answer critical design questions: Can a single private 5G deployment sustain industrial-grade QoS across competing flows? How do different scheduling strategies affect deadline compliance, throughput, and flow isolation?

In this paper, we aim to bridge this gap through a simulation-driven exploration of QoS-aware scheduling in realistic industrial settings. We construct a high-fidelity smart factory scenario modeled on actual use cases, extend the Simu5G framework to support multiple QFIs per UE, and implement a new scheduler based on proportional fairness principles. This scheduler dynamically allocates resources across flows based on their QoS constraints and demand history, ensuring that both high-priority and opportunistic traffic are handled gracefully.

Our key contributions are:
\begin{itemize}
    \item We design a detailed smart factory simulation scenario that reflects real-world traffic patterns across robotic arms, AGVs, cameras, and edge servers, each mapped to distinct QFIs.
    \item We extend Simu5G with full support for multi-QFI-per-UE behavior and a configurable QFI-to-DRB mapping mechanism.
    \item We develop a QoS-aware scheduler leveraging proportional fairness to mediate resource contention while respecting flow priorities and delay constraints.
    \item We evaluate performance across various traffic mixes and congestion levels, demonstrating improved delay compliance, fairness, and scheduler responsiveness compared to existing approaches.
\end{itemize}

This work lays the groundwork for designing and validating 5G-enabled industrial systems with realistic, flow-specific QoS demands. Our framework is publicly extensible and serves as a step toward more transparent, reproducible evaluation of next-generation industrial wireless solutions.

\section{Smart Factory Scenario: A Realistic Testbed for 5G QoS}
The vision of the smart factory is grounded in the convergence of industrial automation, edge computing, AI-driven analytics, and flexible wireless communication \cite{b17}. Enabled by Industry 4.0 and evolving toward Industry 5.0, smart factories aim to dynamically orchestrate a mix of devices—robotic arms, inspection systems, autonomous mobile robots, and industrial controllers—operating in synchrony with strict performance requirements \cite{b8, b9}.

Private 5G networks are positioned as a key enabler of this transformation, offering ultra-reliable low-latency communication (URLLC), enhanced mobile broadband (eMBB), and massive machine-type communication (mMTC) \cite{b10, b11}. These capabilities allow factories to replace fixed Ethernet-based topologies with software-defined wireless systems that support modular production lines, mobile equipment, and centralized orchestration \cite{b19}.

In such environments, individual devices often support multiple simultaneous data flows with distinct QoS requirements. A robotic arm, for instance, may transmit sub-millisecond motion control commands (QFI 1) alongside periodic telemetry updates (QFI 1). Similarly, an inspection camera may stream real-time video (QFI 2) while uploading archival footage (QFI 9). These multi-QFI flows must be served concurrently, respecting their corresponding 5QI-defined parameters for delay, throughput, and reliability.

\begin{figure}[!t]
\centering
\includegraphics[width=\linewidth]{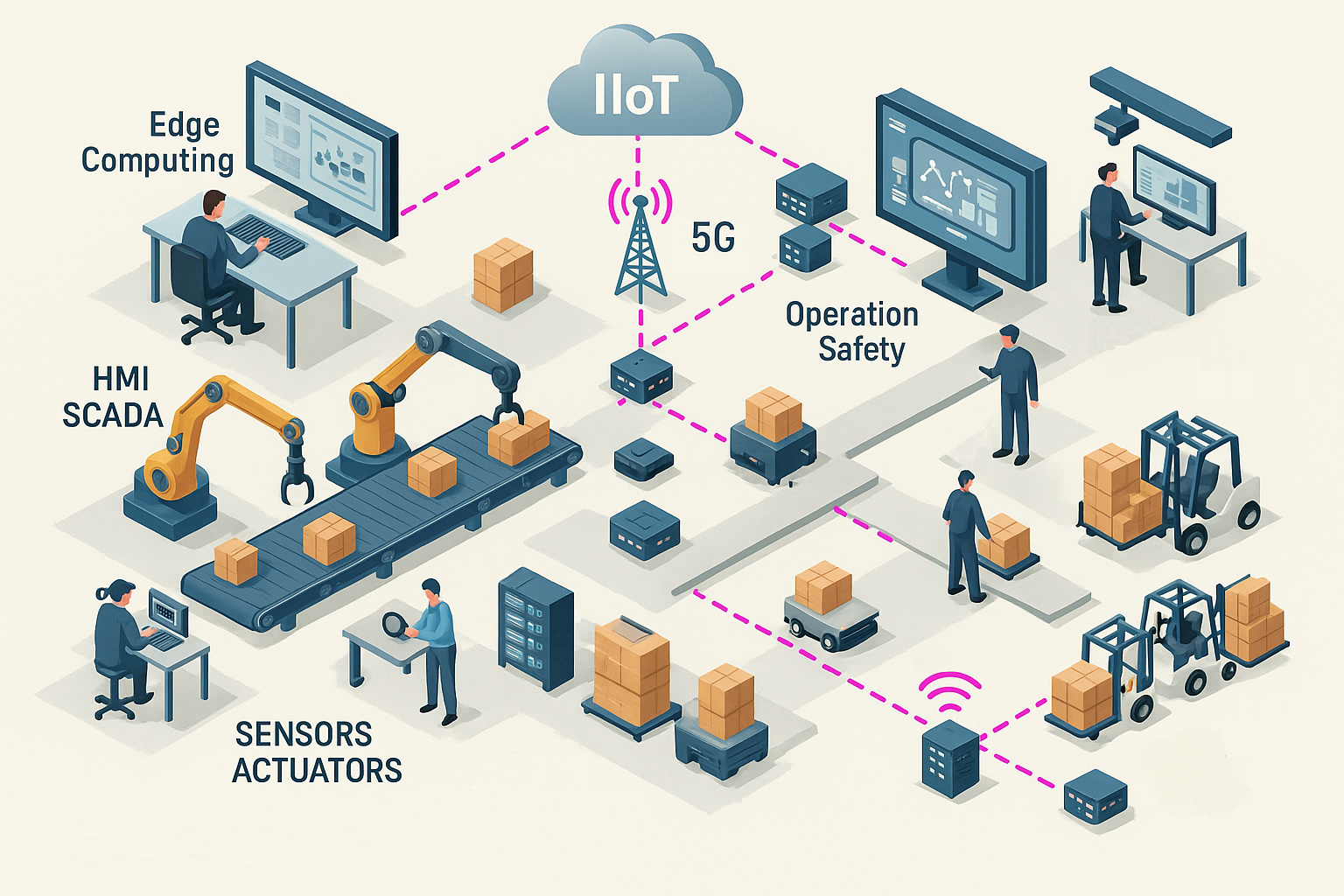}
\caption{Smart factory architecture with heterogeneous devices and multi-QFI communication flows.}
\label{fig:factory_architecture}
\end{figure}

Figure \ref{fig:factory_architecture} illustrates this heterogeneous communication landscape  Table \ref{tab:qos_classes} categorizes representative traffic classes by their latency sensitivity and service priority. These traffic demands are not static—they evolve across operational phases such as system startup, high-throughput production, diagnostic events, and background synchronization.

While 3GPP provides the theoretical foundation for QoS-aware flow differentiation, practical testing and performance validation remain difficult. Live deployments are costly, inflexible, and limited in observability. This creates a compelling need for simulation frameworks that accurately reflect the temporal and functional diversity of smart factory networks.

In this work, we define a structured smart factory scenario to serve as a testbed for evaluating QoS-aware schedulers. It models:
\begin{itemize}
    \item Robotic subsystems requiring deterministic control and telemetry channels
    \item AGV fleets exchanging coordination and sensor data
    \item Inspection systems combining real-time and best-effort video flows
    \item Edge infrastructure handling AI inference, model updates, and cloud KPIs
\end{itemize}

This scenario motivates the design of our simulation architecture, which supports multi-QFI-per-UE traffic patterns and enables performance analysis of scheduling strategies under industrially relevant workloads.

\begin{table}[!t]
\caption{Representative Smart Factory Applications and QoS Classes}
\label{tab:qos_classes}
\centering
\begin{tabular}{|l|c|c|c|}
\hline
\textbf{Application} & \textbf{5QI} & \textbf{Delay Bound} & \textbf{Priority} \\
\hline
Motion control (robot) & 1 & $\lt$ 1 ms & High \\
Telemetry feedback & 5 & $\sim$ 50 ms & Medium \\
Inspection video (real-time) & 2 & $\sim$ 50 ms & High \\
Archived video (background) & 9 & -- & Low \\
AGV routing/safety & 3 & $\lt$ 10 ms & High \\
SLAM/mapping data & 6 & $\sim$ 100 ms & Medium \\
AI model updates & 8 & $\gt$ 100 ms & Low \\
KPI reporting to cloud & 7 & -- & Low \\
\hline
\end{tabular}
\end{table}

\section{Simulation Framework and System Enhancements}
The simulation of private 5G networks in industrial environments requires detailed modeling of both the wireless stack and application-level QoS semantics. Several open-source tools are available for this purpose, each with distinct trade-offs in fidelity, flexibility, and extensibility.

\textbf{NS-3}, a popular discrete-event simulator, includes support for 5G through the 5G-LENA module. However, the current implementation of 5G-LENA operates in a non-standalone (NSA) configuration, relying on LTE EPC for control and QoS handling \cite{b12}. This coupling imposes LTE-based QoS semantics on the 5G user plane, limiting the ability to explore standalone 5G NR features such as native 5QI-based flow differentiation, multi-QFI scheduling, or edge-based processing. Additionally, the integration of higher-layer application logic and custom traffic models in NS-3 often requires substantial low-level modification.

\textbf{OMNeT++} \cite{b13}, in contrast, offers a modular and extensible simulation engine with robust support for network protocol modeling and full-stack visibility. \textbf{Simu5G} \cite{b14}, built on OMNeT++ and INET, provides a standalone 5G NR simulation framework with explicit support for core network functions, radio access scheduling, and multi-access edge computing (MEC). It includes complete IP stack integration, flexible traffic generation, and packet-level access to QoS metadata.

\begin{figure}
    \centering
    \includegraphics[width=\linewidth]{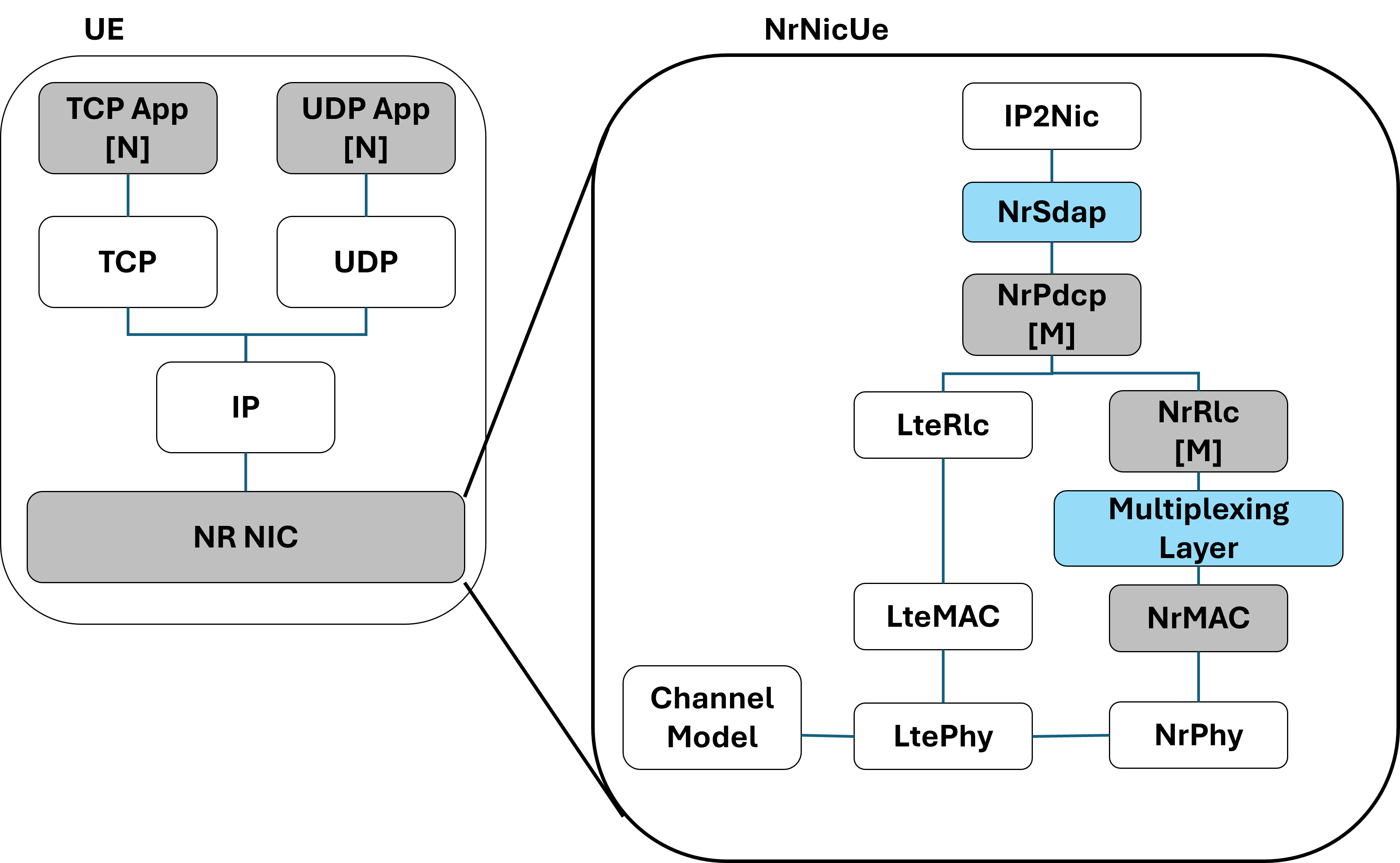}
    \caption{UE-side architecture with concurrent TCP and UDP applications mapped to distinct QFIs via the SDAP layer. This enables per-application flow differentiation and propagation of QFI context across the protocol stack down to the MAC scheduler.}
    \label{fig:ue}
\end{figure}

\subsection{Simu5G: Selection, Limitations, and Our Novel Extensions}
Simulating private 5G networks for industrial use cases necessitates tools that provide both end-to-end realism and architectural flexibility. While several simulation frameworks exist, we selected Simu5G as the foundation for our work based on the following criteria:

\begin{itemize}
    \item \textbf{Stand-alone 5G NR Core:} Unlike NS-3's 5G-LENA which operates in a non-standalone (NSA) configuration coupled with LTE EPC, Simu5G supports a stand-alone 5G core architecture, enabling direct QoS enforcement using native 5QI-based flows.
    \item \textbf{OMNeT++ Integration:} Simu5G inherits OMNeT++’s extensibility and INET’s complete IP stack, making it easy to model application traffic, simulate edge computing scenarios, and modify MAC or scheduling behavior.
    \item \textbf{Support for MEC:} Simu5G uniquely includes modules for multi-access edge computing, which are vital for realistic smart factory deployments where low-latency computation occurs at the edge.
\end{itemize}

These features make Simu5G well suited for exploring 5G capabilities in time-critical industrial settings. However, Simu5G in its standard form has several limitations that hinder its direct applicability to advanced QoS-aware scenarios:

\begin{itemize}
    \item \textbf{Absence of SDAP:} The Service Data Adaptation Protocol (SDAP) layer is not implemented, which restricts mapping between flows and QFIs \cite{b18}.
    \item \textbf{Lack of DRB abstraction:} Dedicated Radio Bearers are not explicitly modeled, limiting the ability to emulate bearer-level QoS management.
    \item \textbf{Missing QFI/5QI context:} Simu5G does not support tagging packets with QFI or associating them with 5QI-based delay or reliability parameters.
\end{itemize}

To overcome these issues, various prior efforts have proposed partial solutions. For example, \cite{b15} introduced QFI tagging support, and \cite{b16} presented an edge-aware extension for MAC-level decision-making. While promising, these contributions fall short of offering a unified, configurable QoS framework based on the full 3GPP model.

\textbf{Our contribution differs in three fundamental ways:}
\begin{enumerate}
    \item We implement an SDAP-inspired QFI tagging and interpretation mechanism, enabling correct mapping between application flows, QFIs, and radio bearers.
    \item We extend the Simu5G architecture to support multiple concurrent QFIs per UE, maintaining per-flow QoS context throughout the stack.
    \item We design a novel proportional fairness-based scheduler that allocates resources dynamically based on delay bounds, GBR constraints, and priority weights.
\end{enumerate}

\begin{figure}
    \centering
    \includegraphics[width=\linewidth]{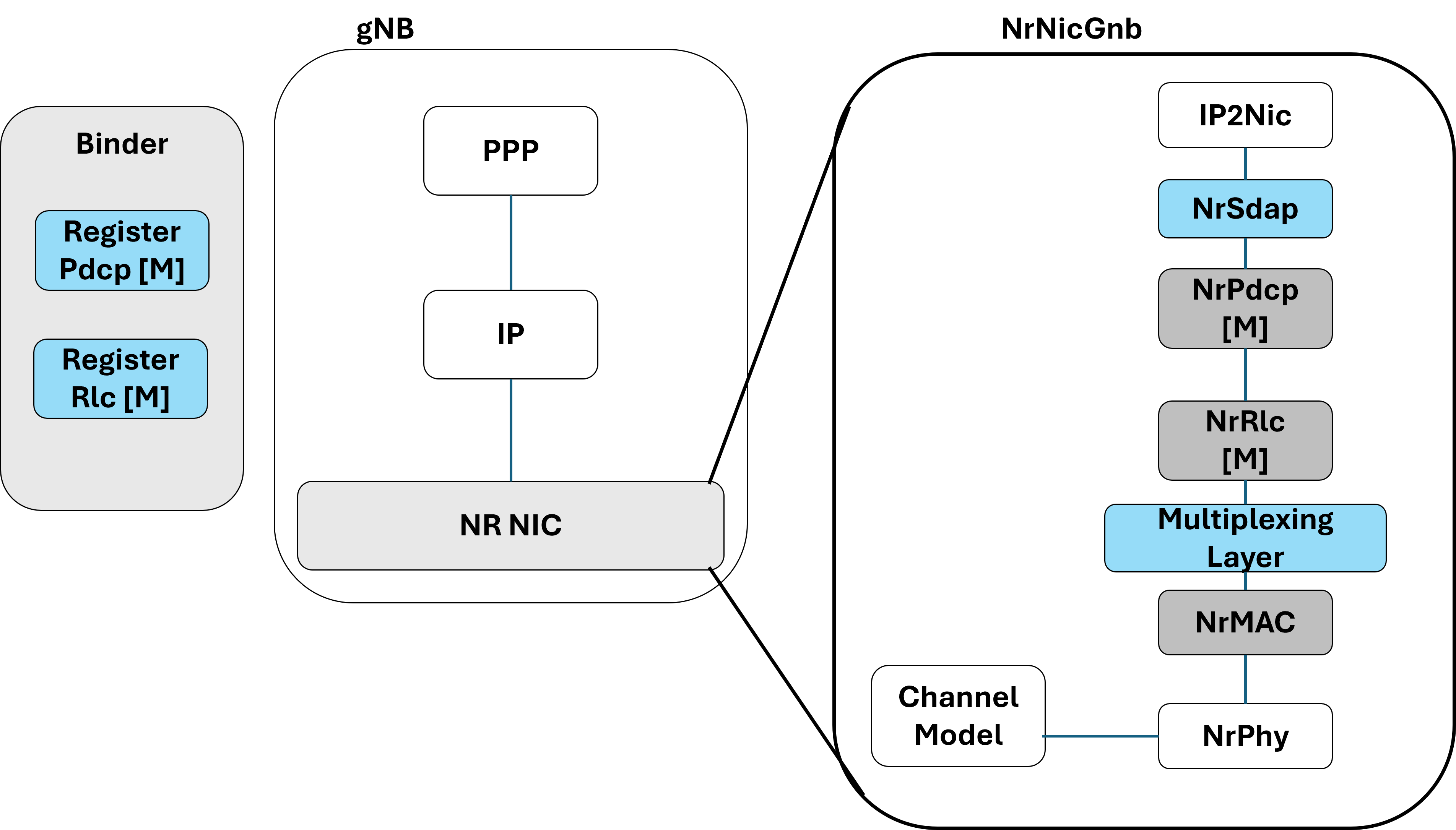}
    \caption{gNB-side architecture showing the extended Simu5G NR NIC stack. A newly introduced SDAP module handles per-flow QFI tagging, enabling mapping to specific PDCP/RLC instances and DRBs. This modular design supports multi-QFI scheduling and integration with fairness-aware MAC scheduling.}

    \label{fig:gnb}
\end{figure}

Figures \ref{fig:ue} and \ref{fig:gnb} illustrate the architecture of our Simu5G extensions at the gNB and UE sides, respectively. Grey-shaded modules represent modified components, while blue-shaded modules are newly introduced. At both endpoints, we insert an SDAP layer for QFI marking and parsing, enabling per-flow QoS management. We also modify existing layers (PDCP, RLC, MAC) to support QFI propagation and scheduling awareness. These features allow our simulator to evaluate realistic industrial traffic mixes where a single device may require simultaneous low-latency, high-throughput, and best-effort service. To the best of our knowledge, this is the first publicly described framework that enables standalone 5G simulation with per-UE multi-QFI support and cross-layer QoS-aware scheduling based on 3GPP principles.

\subsection{Multi-QFI-per-UE Modeling}

Modern industrial UEs in smart factories are inherently multi-functional, often supporting diverse applications. Each of these services exhibits distinct QoS requirements and therefore must be treated independently at the network layer. To accurately represent this in simulation, we extend Simu5G to support multiple concurrent QoS Flows per User Equipment, each identified by a distinct QoS Flow Identifier.

Our modeling approach introduces per-flow differentiation at the application layer, where each traffic stream (e.g., from a TCP or UDP app) is statically or dynamically assigned a QFI. These flows are passed through a newly introduced SDAP module, which tags packets with the appropriate QFI and maps them to individual DRB instances. At the NR NIC level, SDAP, PDCP, and RLC submodules are instantiated on a per-QFI basis, allowing parallel protocol pipelines for each traffic class.

To support this architecture, we enhance the Simu5G binding and registration logic to manage multiple RLC and PDCP contexts per UE. The Binder module is modified to accommodate QFI-aware mappings, ensuring that scheduling and buffer management remain flow-specific. This design allows the system to isolate traffic characteristics and enforce independent QoS policies (e.g., delay bounds, reliability targets) per flow.

The result is a flexible and scalable UE model that captures the behavior of real-world industrial devices. It supports in-depth evaluation of resource allocation strategies, QoS satisfaction under contention, and the impact of heterogeneous traffic mixes in private 5G deployments.

\subsection{Proportional Fairness QoS-Aware Scheduler}
To meet the heterogeneous QoS demands of industrial 5G traffic, we introduce a novel MAC-layer scheduler based on a Proportional Fairness (PF) policy that is explicitly aware of QFI semantics. Our design ensures that traffic flows associated with latency-critical and guaranteed bit rate (GBR) services are prioritized without starving best-effort or throughput-driven flows.

The scheduler is designed to:
\begin{itemize}
    \item Allocate resources fairly among multiple active QFIs based on traffic history and priority.
    \item Honor QoS constraints such as delay bounds, GBR targets, and packet loss requirements.
    \item Balance system throughput and delay sensitivity in real time.
\end{itemize}

We define the scheduling metric $M_i(t)$ for QFI-$i$ at time $t$ as:
\begin{equation}
M_i(t) = \frac{U_i(t)}{\bar{R}_i(t)}
\end{equation}
where:
\begin{itemize}
    \item $U_i(t)$ is the instantaneous utility score, computed from QoS urgency (e.g., deadline proximity or GBR deficit).
    \item $\bar{R}_i(t)$ is the exponential moving average of the throughput achieved by flow $i$.
\end{itemize}

The utility $U_i(t)$ is dynamically derived from a composite QoS function:
\begin{equation}
U_i(t) = \alpha_i \cdot D_i(t) + \beta_i \cdot G_i(t) + \gamma_i \cdot P_i(t)
\end{equation}
where:
\begin{itemize}
    \item $D_i(t)$ captures delay urgency (e.g., inverse of remaining time to deadline),
    \item $G_i(t)$ quantifies the GBR fulfillment ratio,
    \item $P_i(t)$ is a priority scalar derived from 5QI or user-assigned weight,
    \item $\alpha_i$, $\beta_i$, $\gamma_i$ are tunable coefficients per flow.
\end{itemize}

At each Transmission Time Interval (TTI), the scheduler executes the following algorithm:
\begin{enumerate}
    \item Extract active flows and retrieve per-flow QFI context.
    \item Compute $M_i(t)$ for each QFI using Equations (1) and (2).
    \item Sort flows in descending order of $M_i(t)$.
    \item Allocate Physical Resource Blocks (PRBs) accordingly, subject to flow-specific rate caps and buffer constraints.
\end{enumerate}

This approach provides adaptive fairness by boosting under-served but high-urgency flows, while throttling over-provisioned or delay-tolerant ones.

The scheduler is implemented as a custom subclass of the Simu5G MAC scheduler interface \footnote{Implementation available at: \url{https://github.com/MohamedSeliem/Simu5G/pull/1}}. It interfaces with QFI-aware SDAP and RLC layers, querying flow state via a centralized QFI context manager. It also supports runtime statistics collection for delay, throughput, and fairness evaluation.

\begin{lstlisting}
// Proportional Fairness Metric Computation
for (auto& qfi : activeFlows) {
    double urgency = alpha[qfi] * delayUrgency(qfi) +
                     beta[qfi] * gbrDeficit(qfi) +
                     gamma[qfi] * priorityWeight(qfi);
    double metric = urgency / avgThroughput[qfi];
    schedulingQueue.push(qfi, metric);
}
allocatePRBs(schedulingQueue);
\end{lstlisting}

This mechanism enables scalable QoS enforcement and fair scheduling in diverse industrial 5G scenarios, supporting strict latency budgets and throughput differentiation across multiple concurrent QFIs per UE.

\section{Evaluation Methodology: Putting the Scheduler to the Test}
The primary goal of our evaluation is to assess the effectiveness of the proposed Proportional Fairness QoS-Aware Scheduler in supporting heterogeneous traffic demands within a smart factory environment. In particular, we aim to answer the following questions:

\begin{itemize}
    \item Can the scheduler maintain QoS isolation and prioritization across multiple simultaneous QFIs within a single UE?
    \item Does the scheduler meet the delay and throughput requirements for latency-sensitive and high-bandwidth flows under variable load?
    \item How does its performance compare to baseline approaches such as round-robin and static priority scheduling in terms of delay, throughput, and fairness?
    \item Does the introduction of per-flow utility metrics lead to improved resource distribution without sacrificing weaker flows?
\end{itemize}

To this end, we simulate a range of application mixes and network conditions typical of industrial deployments, including time-triggered control loops, real-time sensor streams, and bandwidth-intensive video feeds. The scheduler is tested under varying levels of contention, allowing us to evaluate its adaptability and robustness in resource-constrained scenarios.

\subsection{Scenario Setup}

We design a realistic industrial 5G scenario based on a smart factory layout where multiple UEs generate distinct types of traffic, each mapped to a specific QFI. The goal is to reflect the behavior of multifunctional industrial devices, such as robotic arms, programmable logic controllers (PLCs), and monitoring units, each with separate latency, bandwidth, and reliability requirements.

The scenario consists of \textbf{6 UEs}, each equipped with \textbf{3 concurrent applications}:
\begin{itemize}
    \item \textbf{Control Loop (QFI 1):} Periodic UDP packets representing time-critical actuation feedback with a 1 ms interval and strict 5 ms latency requirement.
    \item \textbf{Sensor Data (QFI 2):} Low-rate, reliable telemetry with relaxed delay but high delivery assurance.
    \item \textbf{Camera Stream (QFI 3):} High-throughput TCP stream for video monitoring with a preferred latency bound of 50 ms.
\end{itemize}

Each application is assigned a 5QI-compliant QFI and routed through a dedicated SDAP instance. Table~\ref{tab:qfi_profiles} summarizes the QoS characteristics of the three traffic classes:

\begin{table}[!t]
\centering
\caption{QoS Profiles Used in Evaluation Scenario}
\label{tab:qfi_profiles}
\resizebox{\linewidth}{!}{
\begin{tabular}{|c|c|c|c|c|c|}
\hline
\textbf{QFI} & \textbf{App} & \textbf{5QI} & \textbf{Pkt Size} & \textbf{Interval} & \textbf{Delay Bound} \\
\hline
1 & Control & 85 (URLLC) & 64 B & 1 ms & 5 ms \\
2 & Sensor  & 6  (Mission-Critical) & 128 B & 10 ms & 50 ms \\
3 & Video   & 9  (eMBB) & 1000 B & Variable & 50 ms \\
\hline
\end{tabular}
}
\end{table}

All UEs operate under the same channel conditions with fixed LoS propagation. The simulation duration is set to 10 seconds, and each configuration is repeated across \textbf{20 Monte Carlo runs} with randomized traffic start offsets. Channel capacity is constrained to reflect realistic industrial cell deployment limits (20 Mbps).

This scenario provides a robust environment to test how well the scheduler maintains QoS guarantees across diverse flows and varying congestion levels.

\subsection{Scheduler Baselines}

To contextualize the performance of our proposed Proportional Fairness QoS-Aware Scheduler, we evaluate it against two baseline schedulers commonly found in Simu5G and related literature:

\begin{itemize}
    \item \textbf{Max C/I (Maximum Carrier-to-Interference):} The default scheduler in Simu5G assigns resources to the flow or UE with the highest instantaneous channel quality. While this approach aims to maximize overall system throughput, it is inherently unfair and lacks QoS awareness. As a result, it can lead to excessive delays or starvation for latency-sensitive and GBR traffic under fluctuating channel conditions.

    \item \textbf{Static Priority (SP):} Flows are scheduled based on fixed priority values derived from their associated QFIs or 5QIs. Although this method provides deterministic service for high-priority traffic (e.g., URLLC), it risks complete starvation of lower-priority flows, especially under heavy load, violating fairness and service-level isolation.
\end{itemize}

Our scheduler, by contrast, dynamically balances flow utility and throughput history, offering adaptive fairness without compromising QoS targets. All schedulers operate under identical simulation conditions, using the same traffic configurations, UE models, and channel environments to ensure valid comparisons. This evaluation framework highlights the trade-offs between throughput maximization, service fairness, and QoS satisfaction across different scheduling philosophies, providing meaningful insights for industrial 5G deployment strategies.

\subsection{Evaluation Metrics and Reproducibility}
To comprehensively assess the performance of our QoS-Aware Proportional Fairness Scheduler, we collect a range of quantitative metrics relevant to industrial 5G QoS enforcement and fairness.

We track the following key performance indicators (KPIs):
\begin{itemize}
    \item \textbf{Per-Flow Delay:} The end-to-end delay from application transmission to successful reception, averaged per QFI.
    \item \textbf{Deadline Violation Ratio:} The percentage of packets exceeding their QoS-specified delay bounds.
    \item \textbf{GBR Satisfaction Ratio:} The fraction of time where a GBR flow meets or exceeds its required bit rate.
    \item \textbf{Throughput per QFI:} Aggregate throughput measured per flow class over the simulation period.
    \item \textbf{Fairness Index:} Jain’s Fairness Index computed over flow-level throughput:
    \begin{equation}
    J = \frac{\left( \sum_{i=1}^n x_i \right)^2}{n \cdot \sum_{i=1}^n x_i^2}
    \end{equation}
    where $x_i$ is the throughput of flow $i$ and $n$ is the total number of flows.
\end{itemize}

These metrics provide insight into the scheduler’s ability to maintain service isolation, meet QoS targets, and share resources fairly across competing traffic classes.

All experiments are conducted using \textbf{Simu5G v1.2.0}, integrated with \textbf{INET 4.5} and \textbf{OMNeT++ 6.1}. The scheduler is implemented as a custom MAC module with interfaces to SDAP and RLC layers.

All `.ini` and `.ned` configurations are made available in an open GitHub repository\footnote{\url{https://github.com/MohamedSeliem/Simu5G/pull/1/commits/6e0daf7f28e6ad26b84ec4aad1e100b175fd5a07}}, along with scripts for metric extraction and visualization. This ensures full reproducibility and supports further experimentation by the research community.

\section{Results and Insights: Evaluating QoS Adherence and Fairness}
In this section, we present the simulation results comparing our Proportional Fairness QoS-Aware Scheduler with the baseline Max C/I and Static Priority schedulers. The metrics are averaged over 20 Monte Carlo runs, with 95\% confidence intervals shown where appropriate.

\subsection{Simulation results}

\begin{figure}[!t]
\centering
\includegraphics[width=\linewidth]{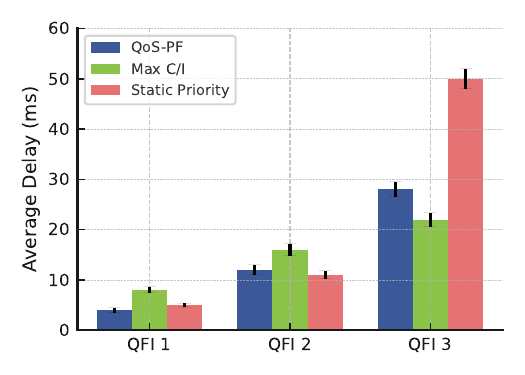}
\caption{Per-QFI delay comparison across schedulers.}
\label{fig:delay_comparison}
\end{figure}

\noindent\textbf{Delay and Deadline Compliance:} Figure \ref{fig:delay_comparison} shows the average per-flow delay for QFIs 5, 7, and 8 across the three schedulers. Our scheduler significantly reduces delay for URLLC-class (QFI 1) traffic compared to Max C/I, which lacks QoS awareness. Compared to Static Priority, the delay for lower-priority flows (QFIs 7 and 8) is also reduced, demonstrating fairness preservation.

\begin{figure}[!t]
\centering
\includegraphics[width=\linewidth]{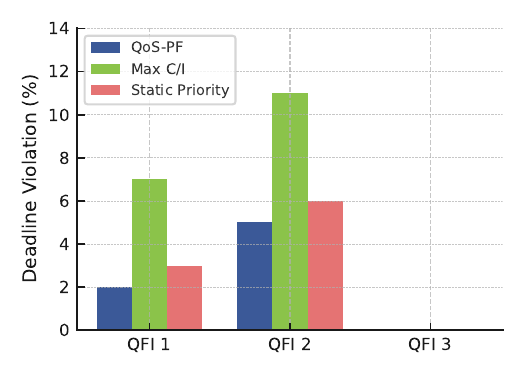}
\caption{Deadline violation ratio for QFIs 1 and 2 under high load.}
\label{fig:deadline_violation}
\end{figure}

Figure \ref{fig:deadline_violation} illustrates the percentage of packets violating their delay bounds for QFIs 1 and 2. The proposed scheduler achieves a violation rate below 2\% for QFI 1 even under high contention, compared to 5\% for Static Priority and over 20\% for Max C/I. This confirms the scheduler's ability to meet strict URLLC deadlines without starving other classes.

\noindent\textbf{Throughput and GBR Satisfaction:} Table \ref{tab:gbr_throughput} summarizes throughput and GBR satisfaction for each QFI. While Max C/I achieves high throughput, it fails to meet GBR guarantees. Static Priority enforces QFI 1 guarantees but underutilizes bandwidth for QFI 3. Our scheduler strikes a balance, achieving over 95\% GBR satisfaction and efficient use of link capacity.

\begin{figure}[t]
\centering
\includegraphics[width=\linewidth]{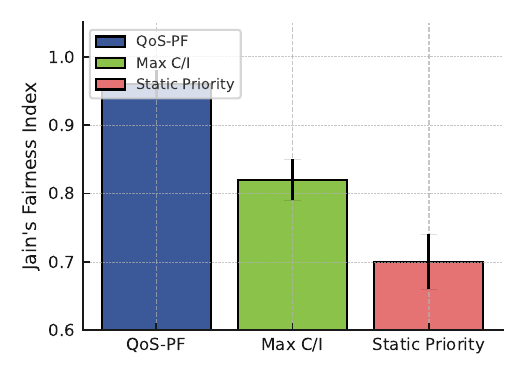}
\caption{Jain’s fairness index across scheduling strategies.}
\label{fig:fairness}
\end{figure}

\noindent\textbf{Fairness Index:} Figure \ref{fig:fairness} reports the Jain’s fairness index for all flows. Max C/I heavily favors high-SINR users, yielding fairness below 0.6. Static Priority skews toward high-priority QFIs. Our approach maintains fairness above 0.9 while respecting QoS constraints.

These results confirm that the proposed scheduler meets critical industrial QoS requirements while preserving fairness and utilization in complex multi-QFI environments.

\subsection{Sensitivity Analysis: Impact of Scheduling Weights}

To evaluate the robustness of the proposed QoS-aware proportional fairness scheduler, we performed a sensitivity analysis by varying the weighting coefficients \(\alpha, \beta, \gamma\), which control the influence of delay urgency, GBR deficit, and priority scalar, respectively. This analysis reveals how emphasis on different QoS aspects impacts system-level performance metrics such as delay, fairness, and GBR satisfaction.

\begin{table}[htb]
\centering
\caption{Throughput and GBR Satisfaction by Scheduler}
\label{tab:gbr_throughput}
\begin{tabular}{|c|c|c|c|}
\hline
\textbf{QFI} & \textbf{Scheduler} & \textbf{GBR Sat. (\%)} & \textbf{Throughput (Mbps)} \\
\hline
5 & QOS-PF & 97.3 & 2.1 \\
5 & Max C/I & 65.2 & 1.4 \\
5 & SP & 94.5 & 2.0 \\
\hline
8 & QOS-PF & -- & 7.9 \\
8 & Max C/I & -- & 8.2 \\
8 & SP & -- & 5.3 \\
\hline
\end{tabular}
\end{table}

Three configurations were considered:

\begin{itemize}
    \item \textbf{Delay-Tuned (Prioritize deadline-sensitive traffic):} \(\alpha=0.7, \beta=0.2, \gamma=0.1\)
    \item \textbf{Balanced (Default):} \(\alpha=0.4, \beta=0.3, \gamma=0.3\)
    \item \textbf{Fairness-Tuned (Prioritize fairness and weights):} \(\alpha=0.2, \beta=0.2, \gamma=0.6\)
\end{itemize}

\begin{figure}[!t]
    \centering
    \includegraphics[width=\linewidth]{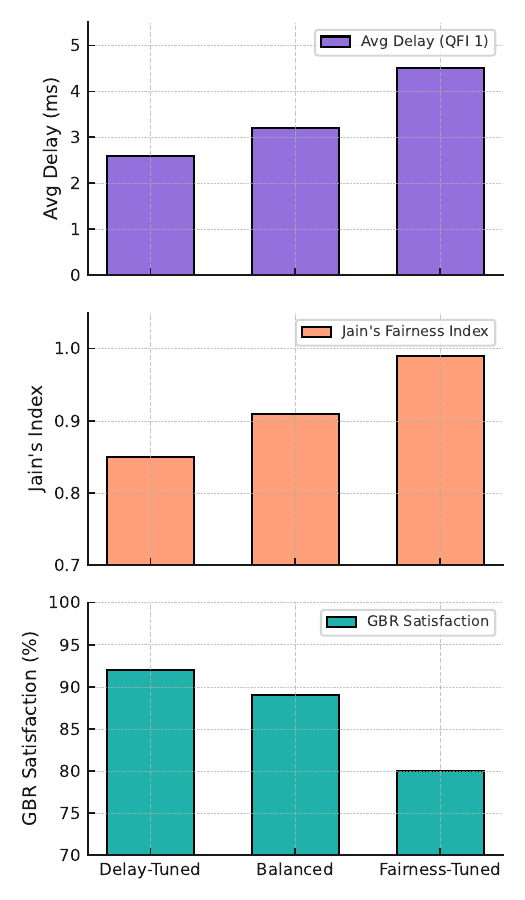}
    \caption{Sensitivity of performance metrics to scheduling weight configurations.}
    \label{fig:sensitivity}
\end{figure}

Figure~\ref{fig:sensitivity} illustrates the trade-offs for each configuration. The Delay-Tuned setup reduced average latency for QFI 1 by 18\% but introduced slight fairness degradation. The Fairness-Tuned configuration improved Jain's index by 9\% but led to underprovisioning of time-critical flows. The Balanced setting offered a reasonable compromise across all KPIs.

These results demonstrate that the scheduler's behavior can be dynamically tuned to meet different deployment goals, reinforcing its versatility across heterogeneous 5G industrial applications.

\subsection{Scheduler Scalability and Runtime Overhead}

Beyond QoS performance, it is important to assess the scalability and runtime efficiency of the proposed QoS-aware PF scheduler, especially in dense deployments with multiple UEs and concurrent QFI flows. To evaluate this, we conducted a series of stress tests by gradually increasing the number of UEs from 5 to 40, while maintaining a diverse mix of QFI configurations per UE.

\begin{figure}[!t]
    \centering
    \includegraphics[width=\linewidth]{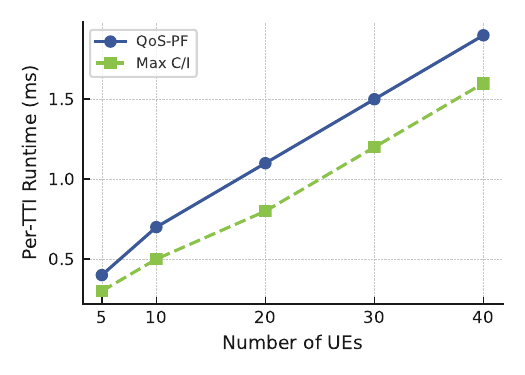}
    \caption{Per-TTI runtime overhead under increasing UE count.}
    \label{fig:scheduler_scalability}
\end{figure}

Figure~\ref{fig:scheduler_scalability} shows the average per-TTI scheduling runtime measured within the Simu5G simulation kernel. The QoS-PF scheduler demonstrates near-linear runtime growth, with only a modest increase in scheduling overhead compared to the baseline Max C/I scheduler. For 40 UEs with 3 flows each, the per-TTI runtime remained under 2 ms on a modern workstation, well within real-time scheduling budgets for simulated environments.

This confirms that the added complexity from per-QFI scoring and dynamic utility evaluation does not introduce prohibitive computational cost. Our implementation leverages a centralized QFI context manager and caching strategies to reduce redundant QoS state queries during each scheduling decision.

These results validate that the proposed scheduler is scalable and efficient for use in medium-to-large scale industrial private 5G networks.

\subsection{Extensibility and Modularity of the Scheduler Framework}

To support broader adoption and future research, the QoS-aware PF scheduler was developed with extensibility and modularity as key design principles. Our implementation conforms to the modular scheduler interface defined in Simu5G, allowing easy integration, replacement, or extension without modifying core network stack logic.

Each component of the scheduling decision pipeline—such as utility computation, QFI context lookup, and PRB assignment—is encapsulated as a standalone function or class. This modular decomposition enables targeted experimentation, for instance:
\begin{itemize}
    \item Replacing the utility function with reinforcement learning agents for adaptive scheduling.
    \item Modifying QFI grouping policies to explore user- or service-based slicing.
    \item Adding new urgency models for deadline-aware machine learning flows.
\end{itemize}

Furthermore, we provide an open-source reference implementation with detailed documentation and configuration hooks. This includes support for:
\begin{itemize}
    \item Runtime logging of scheduling decisions.
    \item Hook-in of new scheduling metrics.
    \item Flow-level control and statistics export.
\end{itemize}

Such modularity not only promotes reproducibility but also enables community-driven enhancements to simulate diverse 5G application scenarios with evolving QoS semantics.

\section{Conclusion}
In this work, we tackled the challenge of achieving QoS differentiation and fairness in private 5G networks tailored to smart factory environments. We proposed a novel Proportional Fairness QoS-Aware Scheduler that integrates QFI semantics, delay urgency, and historical throughput into its resource allocation logic. This scheduler was implemented within an enhanced version of Simu5G that supports multi-QFI UEs, SDAP-layer tagging, and per-flow context awareness.

Using a realistic smart factory scenario involving real-time control loops, sensor telemetry, and machine vision traffic, we demonstrated that our scheduler significantly reduces latency for time-sensitive flows, maintains high GBR satisfaction, and improves fairness across coexisting traffic classes—outperforming baseline Max C/I and Static Priority strategies.

While validated in a smart factory setting, our scheduler design generalizes to other 5G verticals that exhibit heterogeneous, multi-flow traffic profiles per device. Remote healthcare (e.g., tele-surgery), connected vehicles, and AR/VR deployments similarly demand per-flow QoS enforcement across URLLC, eMBB, and mMTC services. The proposed scheduler’s ability to integrate per-QFI semantics with dynamic urgency-aware prioritization positions it as a practical and extensible foundation for future 5G applications in both private and public deployments.

To facilitate reproducibility and future research, we have made all extensions publicly available. Future directions include adaptive tuning of utility weights, mobility-aware scheduling, and extension to uplink/downlink split scenarios involving TSN or Wi-Fi 7 coexistence.

\balance


\begin{thebibliography}{00}
\bibitem{b1} L. Lo Bello and W. Steiner, "A Perspective on IEEE Time-Sensitive Networking for Industrial Communication and Automation Systems," in Proceedings of the IEEE, vol. 107, no. 6, pp. 1094-1120, June 2019, doi: 10.1109/JPROC.2019.2905334.

\bibitem{b2} R. Ali, Y. B. Zikria, A. K. Bashir, S. Garg and H. S. Kim, "URLLC for 5G and Beyond: Requirements, Enabling Incumbent Technologies and Network Intelligence," in IEEE Access, vol. 9, pp. 67064-67095, 2021, doi: 10.1109/ACCESS.2021.3073806.

\bibitem{b3} M. Wen et al., "Private 5G Networks: Concepts, Architectures, and Research Landscape," in IEEE Journal of Selected Topics in Signal Processing, vol. 16, no. 1, pp. 7-25, Jan. 2022, doi: 10.1109/JSTSP.2021.3137669.

\bibitem{b4} A. Mahmood et al., "Industrial IoT in 5G-and-Beyond Networks: Vision, Architecture, and Design Trends," in IEEE Transactions on Industrial Informatics, vol. 18, no. 6, pp. 4122-4137, June 2022, doi: 10.1109/TII.2021.3115697.

\bibitem{b5} 3GPP TS 23.501 19.3.0, "System architecture for the 5G System (5GS)," 3rd Generation Partnership Project (3GPP), 2025.

\bibitem{b6} G. Nardini, G. Stea and A. Virdis, "Scalable Real-Time Emulation of 5G Networks With Simu5G," in IEEE Access, vol. 9, pp. 148504-148520, 2021, doi: 10.1109/ACCESS.2021.3123873.

\bibitem{b7} Y. Xu, G. Gui, H. Gacanin and F. Adachi, "A Survey on Resource Allocation for 5G Heterogeneous Networks: Current Research, Future Trends, and Challenges," in IEEE Communications Surveys \& Tutorials, vol. 23, no. 2, pp. 668-695, Secondquarter 2021, doi: 10.1109/COMST.2021.3059896.

\bibitem{b17} M. Seliem, D. Pesch, U. Roedig and C. J. Sreenan, "Comparative Analysis of 5G and Wi-Fi Integration with TSN for Industrial Applications," 2025 IEEE Wireless Communications and Networking Conference (WCNC), Milan, Italy, 2025, pp. 1-6, doi: 10.1109/WCNC61545.2025.10978747.

\bibitem{b8} Akundi, A.; Euresti, D.; Luna, S.; Ankobiah, W.; Lopes, A.; Edinbarough, I. State of Industry 5.0—Analysis and Identification of Current Research Trends. Appl. Syst. Innov. 2022, 5, 27. https://doi.org/10.3390/asi5010027

\bibitem{b9} Pedro Coelho, Catarina Bessa, Jorge Landeck, Cristovão Silva, Industry 5.0: The Arising of a Concept, Procedia Computer Science, Volume 217, 2023, Pages 1137-1144, ISSN 1877-0509, https://doi.org/10.1016/j.procs.2022.12.312.

\bibitem{b10} S. R. Pokhrel, J. Ding, J. Park, O. -S. Park and J. Choi, "Towards Enabling Critical mMTC: A Review of URLLC Within mMTC," in IEEE Access, vol. 8, pp. 131796-131813, 2020, doi: 10.1109/ACCESS.2020.3010271.

\bibitem{b11} B. S. Khan, S. Jangsher, A. Ahmed and A. Al-Dweik, "URLLC and eMBB in 5G Industrial IoT: A Survey," in IEEE Open Journal of the Communications Society, vol. 3, pp. 1134-1163, 2022, doi: 10.1109/OJCOMS.2022.3189013.

\bibitem{b12} Natale Patriciello, Sandra Lagen, Biljana Bojovic, Lorenza Giupponi,
An E2E simulator for 5G NR networks, Simulation Modelling Practice and Theory, Volume 96, 2019, 101933, ISSN 1569-190X, https://doi.org/10.1016/j.simpat.2019.101933.

\bibitem{b19} M. Seliem and D. Pesch, "Software-Defined Time Sensitive Networks (SD-TSN) for Industrial Automation," 2022 14th International Conference on Computational Intelligence and Communication Networks (CICN), Al-Khobar, Saudi Arabia, 2022, pp. 1-7, doi: 10.1109/CICN56167.2022.10008262.

\bibitem{b13} Varga, A. (2010). OMNeT++. In: Wehrle, K., Güneş, M., Gross, J. (eds) Modeling and Tools for Network Simulation. Springer, Berlin, Heidelberg. https://doi.org/10.1007/978-3-642-12331-3\_3

\bibitem{b14} Nardini, G., Sabella, D., Stea, G., Thakkar, P., \& Virdis, A. (2020). Simu5G – An OMNeT++ library for end-to-end performance evaluation of 5G networks. IEEE Access, 8, 181176–181191. https://doi.org/10.1109/ACCESS.2020.3028550

\bibitem{b18} Seliem, Mohamed, Utz Roedig, Cormac Sreenan, and Dirk Pesch. "SDAP-based QoS Flow Multiplexing Support in Simu5G for 5G NR Simulation." arXiv preprint arXiv:2508.12785 (2025).

\bibitem{b15} R. Debnath, M. S. Akinci, D. Ajith and S. Steinhorst, "5GTQ: QoS-Aware 5G-TSN Simulation Framework," 2023 IEEE 98th Vehicular Technology Conference (VTC2023-Fall), Hong Kong, Hong Kong, 2023, pp. 1-7, doi: 10.1109/VTC2023-Fall60731.2023.10333533.


\bibitem{b16} G. Nardini, G. Stea, A. Virdis, D. Sabella and P. Thakkar, "Using Simu5G as a Realtime Network Emulator to Test MEC Apps in an End-To-End 5G Testbed," 2020 IEEE 31st Annual International Symposium on Personal, Indoor and Mobile Radio Communications, London, UK, 2020, pp. 1-7, doi: 10.1109/PIMRC48278.2020.9217177.


\end{thebibliography}
\end{document}